# Low threshold, room-temperature microdisk lasers in the blue spectral range

Igor Aharonovich*,(a)[1], Alexander Woolf*[1], Kasey J. Russell[1]*, Tongtong Zhu[2], Menno J. Kappers[2], Rachel A. Oliver[2] and Evelyn L. Hu[1]

* these authors contributed equally to this work

(a) igor@seas.harvard.edu

1. School of Engineering and Applied Sciences, Harvard University, Cambridge, Massachusetts, 02138, USA

2. Department of Materials Science and Metallurgy, University of Cambridge, Pembroke Street, Cambridge CB2 3QZ, United Kingdom

Abstract

**InGaN-based active layers within microcavity resonators offer the potential of low threshold lasers in the blue spectral range. Here we demonstrate optically pumped, room temperature lasing in high quality factor GaN microdisk cavities containing InGaN quantum dots (QDs) with thresholds as low as 0.28 mJ/cm$^2$. This work, the first demonstration of lasing action from GaN microdisk cavities with QDs in the active layer, provides a critical step for the nitrides in realizing low threshold photonic devices with efficient coupling between QDs and an optical cavity.**

High quality factor ($Q$), low volume microcavities in GaN-based materials offer the prospect of low threshold lasers in the UV to visible range[1-6]. Quantum well (QW) active regions have been incorporated into microdisks[3,6], microrings[7], and photonic crystal structures[8,9]. In general, the room-temperature thresholds of the structures have been in the range of a few mJ/cm$^2$ (for pulsed measurements)[10], or a few hundred kW/cm$^2$ (CW)[6,11]. This work describes the incorporation of quantum dot (QD) – containing active regions within high $Q$ (~ 6,000) microdisk cavities. Lasing was achieved for 1 micron diameter microdisks incorporating 3 layers of QDs, at threshold energies as low as *0.2 mJ/cm$^2$*. Our experiments took place with the concurrent evaluation of four sample structures which allowed us to probe the influence of QD density and cavity design on the lasing threshold.

A schematic illustration of the sample structure is shown in Figure 1a and Figure 1b summarizes the properties of the four materials used in this work. Briefly, the top disk membrane has a thickness of either 120 nm or 200 nm and encapsulates either one or three layers of In$_{0.2}$Ga$_{0.8}$N with each layer containing QDs with an approximate areal density of 1 × 10$^{10}$ cm$^{-2}$ based on atomic force microscopy studies of QD epilayers grown contemporaneously[12]. It should be noted that the InGaN layers contain both QDs and an inhomogeneous quantum well (QW). The implications of this structure will be discussed later.

The pedestal of the disk is an $In_xGa_{1-x}N/ In_yGa_{1-y}N$ sacrificial superlattice (SSL, x = 5.1%, y = 6.5% In) which is grown on an n-doped GaN pseudo-substrate and is capped by a thin ~ 10 nm GaN layer, followed by an $Al_{0.2}Ga_{0.8}N$ etch stop layer (see Methods for more details)[12,13]. The samples have a similar surface pit density, indicating that they have a similar threading dislocation density. This allows us to make the first-order assumption that variations in the performance of the cavities or the lasers are not due to variations in principal defect densities in the starting materials.

Microdisk cavities were formed concurrently from all four material samples, using techniques that have been previously described[11,14]. Photoelectrochemical (PEC) etching of the sacrificial superlattice was used to form the pillars of the microdisks. Figure 1c summarizes the procedure to fabricate the microdisks (see Methods for more details). Figure 1d shows a scanning electron microscope micrograph of a fabricated microdisk from sample A.

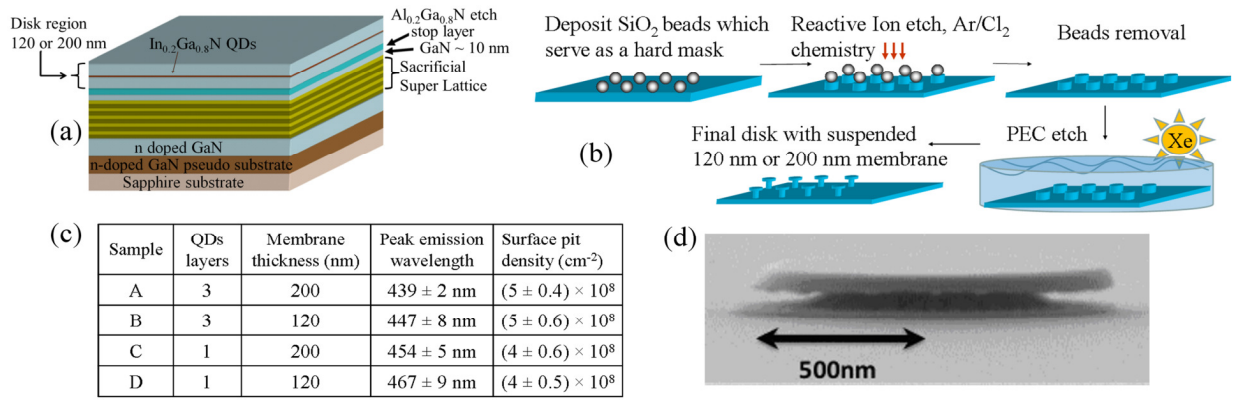

Figure 1. *(a)* A schematic illustration of the sample structure. The disk region has a thickness of 120 or 200 nm with either one or three embedded QD layers. *(b)* Properties of the four investigated samples. *(c)* Schematic illustration of the procedure used to fabricate the microdisks. *(d)* Scanning electron microscope images (top and side view) of the 1 μm diameter microdisk cavity.

The optical properties of the microdisk resonators were investigated using a frequency doubled, pulsed, Titanium-Sapphire laser emitting at 380 nm (76 MHz repetition rate, 200 ps pulse duration) through a high (0.95) numerical aperture (N.A.) objective normal to the surface of the microdisk. The luminescence from the microdisks was collected through the same objective. Photoluminescence (PL) spectra recorded at room temperature from 1 μm diameter microdisks show whispering gallery modes (WGMs) modulating the broad emission from the QD-containing layers (Figure 2a). Figure 2b shows a high resolution spectrum of the transverse electric ($TE_{1,31}$) mode at ~ 477 nm from sample A. This mode exhibits a splitting between the two normally-degenerate counter-propagating whispering gallery modes that is likely due to slight imperfections in fabrication that destroy the rotational symmetry of the disk. These two modes are fitted with Lorentzian functions, as indicated with red lines in Fig 2b. The Q values for each resonance were determined by calculating $\lambda_{cav}/\Delta\lambda_{cav}$, where $\lambda_{cav}$ is the cavity mode wavelength and $\Delta\lambda_{cav}$ is the full width at half maximum (FWHM) of the mode. The highest measured $Q$ for 1μm-diameter disks from sample A, $Q \sim 6600$, is one of the highest $Q$s reported from GaN – based microdisks[15].

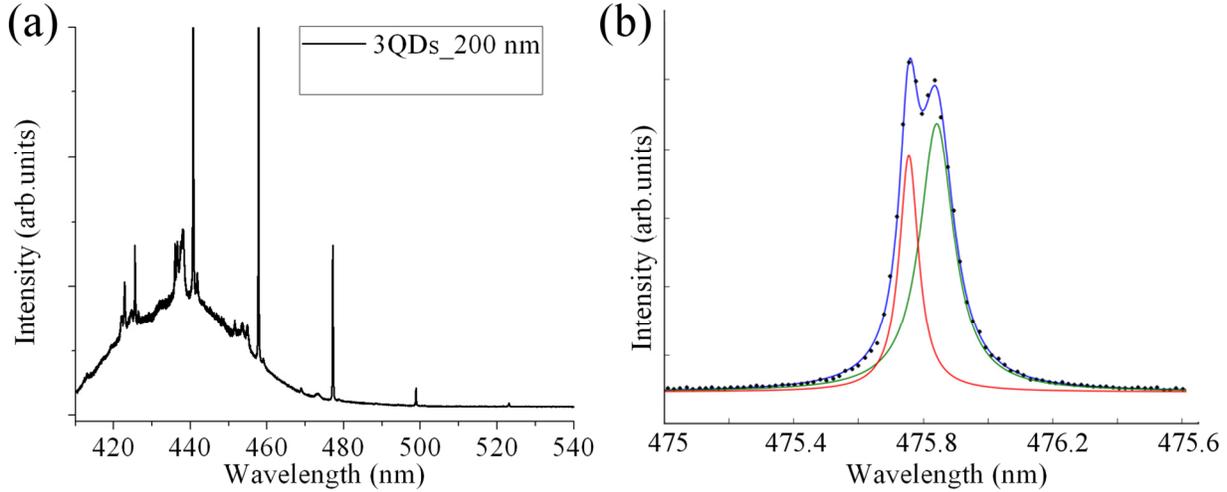

*Figure 2. Optical characterization of a microdisk made of material A. **a.** PL spectrum recorded using 380 nm excitation wavelength, showing predominantly first order WGMs decorating the broad QD –related emission. **b**. High resolution PL spectrum of a high Q mode with $Q \sim 6600$. Green and red lines are fit to a Lorentzian.*

Lasing behavior in microdisks is observed through the dependence of the PL emission intensity and mode linewidth on excitation power, as is shown in Fig. 3a. At excitation powers below the lasing threshold, multiple WGMs are observed. As the excitation power is increased through the lasing threshold, the intensity of a single mode increases abruptly and at higher powers dominates the emission. This behavior is indicative of a transition from spontaneous emission to lasing. The lasing mode in this case was determined to be the $TE_{1,13}$ mode based on FDTD simulations. A plot of the output intensity versus input power for microdisks from sample A (blue triangles) and sample B (red circles) is shown in Fig. 3b, with clear thresholds at ~ 0.28 mJ/cm$^2$ and 0.63 mJ/cm$^2$, respectively. The lasing threshold was determined as the intersection of the horizontal axis and a linear fit to the higher-power region of the data. The same data plotted on a log-log scale (top right inset, Fig. 3c) clearly shows all three regimes of operation: spontaneous emission, amplified spontaneous emission, and laser oscillation[16]. In addition, we observed a pronounced narrowing of the lasing mode as the excitation power was increased through the lasing threshold, signifying the increased temporal coherence of emission in the lasing regime (Fig. 3c). Taken together, these data unambiguously demonstrate the achievement of lasing behavior in our devices. The bottom left inset of Fig 3c shows the optical image of the microdisk laser above lasing threshold recorded using a CCD camera.

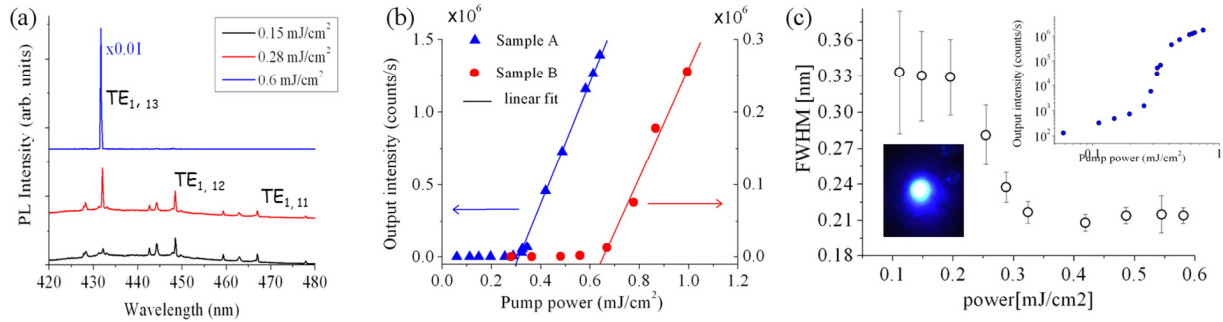

*Figure 3. GaN microdisk lasers. **a.** PL spectra as a function of excitation power recorded from a 1 μm size microdisk cavity (Sample A). **b.** Optical output power of the microdisk laser as a function of excitation power for sample A (blue triangles) and B (red circles). A clear lasing threshold is observed at 0.28 mJ/cm² and 0.63 mJ/cm² for sample A and sample B, respectively. The lines are a linear fit to the data above threshold. **c.** Llinewidth of the lasing mode plotted as a function of excitation power for sample A. The reduction of the linewidth is in accord with lasing behavior. Inset top right, same data as in (b) replotted on a logarithmic scale. Inset bottom left, an optical image of the microdisk laser above threshold recorded using a CCD camera.*

A comparison among samples A-D provides an indication of the most critical factors affecting the lasing behavior of these devices. All microdisks included in this analysis exhibited pronounced modes in the PL spectra with a minimum $Q$ of 1000. Of the four samples investigated in this study, only sample A and sample B exhibited lasing behavior. Sample A showed lower threshold lasing because it exhibits both high $Q$ and it has three layers of QDs that contribute to the emission gain. The other samples either have thinner membranes, which finite difference time domain (FDTD) simulations suggest should lead to lower maximum $Q$ (~18,000) for 120 nm membrane versus (25,000) for the 200 nm membrane), or have fewer layers of QDs. These variations in theoretical and actually-observed $Q$'s of the structure and in density of the gain medium provide an excellent probe of the critical parameters in material and structure that give rise to low threshold lasing.

For all disks that achieved lasing, values were extracted for the threshold power, mode wavelength, and maximum cavity $Q$ measured at low excitation power (Fig. 4). On sample A, 6 of 10 disks achieved lasing, with a threshold that varied from 0.28 – 0.92 mJ/cm². Sample B, which had the same number of layers of QDs as sample A but a thinner membrane, achieved lasing in only 3 of 10 disks, with thresholds in the range 0.4 – 1.1 mJ/cm². The typical $Q$ of modes from microdisks on sample B was lower than on sample A (Fig 4a). This trend is consistent with FDTD simulations, although in both samples the measured $Q$ is more than an order of magnitude lower than the theoretical limit. This latter is not unusual, since the simulations do not account for imperfections in the material or in the fabrication of the microdisks.

In samples C and D, which only contained a single layer of QDs, lasing was not observed on any disks under the excitation powers available in our experimental setup. In addition, the $Q$ of these cavities were comparable to or lower than cavities from samples A and B, respectively. These observations suggest that re-absorption within the InGaN layers themselves is not the dominant factor limiting $Q$ in these structures and that, at least under pulsed excitation at room temperature, multiple QD layers are necessary to supply sufficient gain to compensate losses. The importance of the areal density of quantum dots, and how it limits the modal gain of the active layer has been earlier noted for InGaAs-based quantum dot lasers[17]. Indeed, successful larger-cavity, in-plane lasers with InGaN quantum dot active material have employed 8 to 10 QD layers[5,18,19]. We also note that no clear dependence is observed between the lasing threshold and the $Q$ of the microdisk. Low InGaN re-absorption is consistent with the low number of QDs coupled to a mode: each 1 µm-diameter disk contains approximately 100 QDs per layer at a QD density of ~$10^{10}$ cm$^{-2}$, and of these, it is likely that fewer than 10 are in spatial and spectral resonance with a WGM.

Because re-absorption is sufficiently low in these devices, modes are visible across the entire gain spectrum and not only at the low-energy side of the spectrum (as is typical for GaN microdisks with quantum well active region[3,20]). In fact, in our devices the lasing mode was consistently located at wavelengths shorter than 430 nm, on the high-energy side of the broad QD emission spectrum (Fig. 4b). This dramatic blue-shift of the lasing wavelength for pulsed, rather than CW optical excitation has been observed previously in InGaAs QD microdisk lasers[21]. The reasons for this behavior will be explored further, but may relate to a differential change in radiative lifetimes within a cavity environment with a high instantaneous charged carrier background. The complex interplay between the energies of the lasing mode and the QD gain spectrum is evident in the variation of lasing threshold with wavelength (Fig. 4b). The threshold decreases at longer wavelengths as the overlap with the QD emission spectrum increases. Nevertheless, lasing from a mode at the center of the QD emission spectrum was not observed.

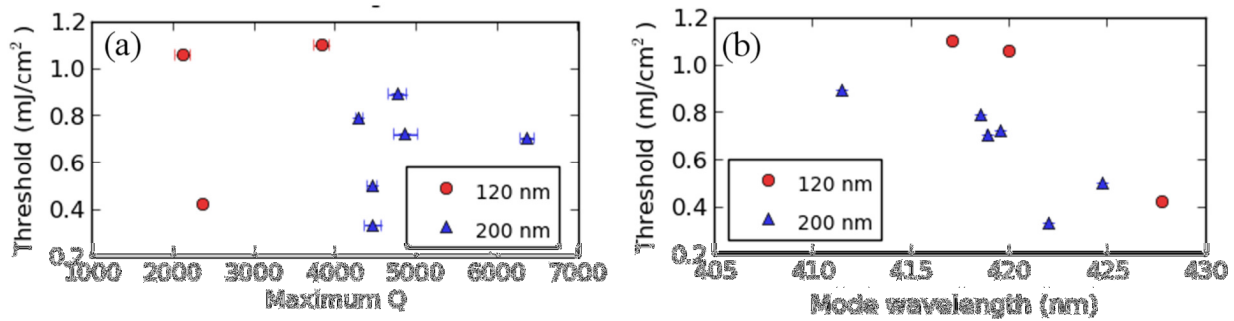

Figure 4. Statistical characterization of the GaN microdisk lasers. *a.* Lasing threshold is plotted as a function of maximum Q recorded from the same microdisk for sample A (blue triangles) and sample B

*(red circles). **b.** Lasing threshold is plotted as a function of a lasing wavelength for sample A (blue triangles) and sample B (red circles).*

Several aspects of this comparison highlight the need for further understanding of the interplay of active layer composition and cavity structure in determining the lasing thresholds of the resulting devices. Further exploration is needed on the optimal InGaN QD material structure , the influence of background defects in the materials, the wavelength-dependence of the QD radiative lifetimes and efficiencies, as well as optimal cavity designs. In addition, our gain medium consists of InGaN layers which contain not only QDs but also an inhomogeneous ('patchy') quantum well layer, with AFM data indicating that the QDs may be located not only on top of the QW layer (similar to the case of Stranski-Krastanov growth in, for example InAs/GaAs, where QDs sit on top a wetting layer) but also in the 'patches' between quantum well regions. Hence, the luminescence peaks from the QDs and the QW overlap, and whilst QD emission has been confirmed from these samples using low temperature microphotoluminescence[12] the influence of the QD and the QW elements of the active region on the room temperature properties cannot be easily separated.

To summarize, we have characterized a set of microdisk lasers with quantum dot-containing active layers. Of the four samples investigated in this study, only sample A and sample B – containing 3 layers of QDs exhibited lasing behavior with thresholds as low as 0.28 mJ/cm$^2$ under pulsed excitation at room temperature. Our results suggest the critical importance of sufficiently high QD areal density. Although the lasing thresholds of these devices are exceptionally low, the sparser areal density of the quantum dot gain material, compared to quantum wells, may explain the yet lower threshold of 300 W/cm$^2$ reported by Tamboli et al[3] for a QW microdisk laser under CW excitation at room temperature. A better strategic design of the InGaN QD active layer material, matched to a lower mode volume cavity (e.g. a photonic crystal cavity) may result in still lower values of lasing thresholds. We believe that our studies are important not only for the efficient lasing performance demonstrated, but also because of the important insights gained on the relative impact on lasing of the materials composition and structure, matched to the microdisk design and fabrication.

**Methods**

**Sample growth**

All samples were grown by metalorganic vapor phase epitaxy (MOVPE) on *c*-plane GaN pseudo-substrates using sapphire substrates in a 6 x 2" Thomas Swan close-coupled showerhead reactor. The pseudo-substrate growth method is described in more detail elsewhere[22]. Trimethylgallium, trimethylindium, trimethylaluminium, silane and ammonia were used as precursors, with $N_2$ used as a carrier gas for the growth of all indium-containing layers and $H_2$ carrier gas for all other layers, except for the GaN layers grown between the SSL and the AlGaN layer, the GaN barrier layers between the QD layers and the first 10 nm of the GaN capping layer on top of the final QD layer. These GaN layers were grown under $N_2$ in order to protect the surface of the SSL and the QD layers from being exposed to and degraded by $H_2$. The InGaN QD layer was grown using a self-assembled droplet epitaxy method which is described in more detail elsewhere[13] using optimized barrier growth methodologies[12].

**Device fabrication**

A schematic illustration of the microdisk fabrication is shown in Fig 1. The microdisks were fabricated by dispersing 1 μm silica beads onto the samples which served as etch hard masks. The samples were then etched using a 500 W inductively coupled plasma reactive ion etching process in an argon/chlorine environment, with a flow rate of 25/25 standard cubic centimeters per minute, respectively. The silica beads were removed by immersing the samples in water in an ultrasonic bath for five minutes. A metal film (Ti/Au), which serves as the cathode for the PEC etching, was then deposited using an e-beam evaporator. Finally, to optically isolate the membrane layer, the microdisks were undercut using PEC etching. The disks were immersed in 0.004M HCl solution and illuminated using a 1000W Xe lamp for 18 minutes. A GaN filter was used to generate electron-hole pairs only in the sacrificial post pedestal layer (below the GaN bandgap), thus etching it away and creating the undercut.

**Optical measurements**

The optical measurements were carried out at room temperature using a home-built confocal microscope system. A frequency-doubled titanium sapphire laser emitting at 380 nm with a pulse repetition of 76 MHz was used as an excitation source. The laser beam was incident on the top of the microdisks through a high N.A objective (×100, numerical aperture = 0.95), creating a ~ 500 nm optical spot. The quoted power was measured underneath the objective in the focal plane of the sample. The emission from the microdisk was collected through the same objective and directed into a spectrometer (Princeton instruments, 1200 lines/mm) through a long pass 400 nm filter, a lens and a 10 μm single mode fiber to achieve confocality.

**Calculations and simulations**

2D FDTD simulations were performed to identify the modal volume, the theoretical quality factor and the WGMs of the microdisks. Index of refraction of n=2.48 was used in the calculations. The spacing of the WGM at the microdisks is given by $\Delta\lambda = \lambda^2/2\pi r n_{eff}$, where $\lambda$ is the mode wavelength, $\Delta\lambda$ is the spacing between adjacent modes, r is the disk radius and $n_{eff}$ is the effective index of refraction. For the 1 µm disks, the spacing is approximately ~ 20 nm.

**Acknowledgments**

The authors acknowledge Haitham A.R. El-Ella for assistance with the sample growth. This work was supported in part by the Engineering and Physical Sciences Research Council (Award No. EP/H047816/1) and by the NSF Materials World Network (Award No. 1008480). This work was enabled by facilities available at the Center for Nanoscale Systems, a member of the National Nanotechnology Infrastructure Network (NNIN) which is supported by the National Science Foundation under NSF award no. ECS-0335765.